\documentclass[a4paper,10pt]{article}
\usepackage[utf8]{inputenc}
\usepackage[T1]{fontenc}
\usepackage{lmodern}
\usepackage{a4wide}
\usepackage{amsmath,amsfonts,amssymb}
\usepackage{bbm}
\usepackage[affil-it]{authblk}
\usepackage{color}
\usepackage[labelfont=bf]{caption}
\usepackage{graphicx}
\usepackage{floatrow}
\usepackage{caption}
\usepackage{subcaption}
\usepackage{soul}

\title{Rare event simulation in immune biology:
Contrasting models of negative selection in T-cell maturation}

\newcommand{\gact}{g_\textrm{act}}
\newcommand{\Tau}{\mathcal{T}}
\newcommand{\Wij}{W_{ij}}
\newcommand{\Tij}{\mathcal{T}_{ij}}
\newcommand{\gthy}{g_\textrm{thy}}

\title{Rare event simulation in immune biology:
Models of negative selection in T-cell maturation}
\author[1]{Corinna Ernst}
\author[2]{Ellen Baake}
\affil[1]{Genome Informatics, Institute of Human Genetics, University of Duisburg-Essen}
\affil[2]{Biomathematics \& Theoretical Bioinformatics, Faculty of Technology, Bielefeld University}

\date{}

\begin{document}

\maketitle 
\thispagestyle{empty}

\begin{abstract}
We present an application of rare event simulation in the area
of immune biology.
A major task of the immune system is the recognition of foreign antigens, which enter the body as parts of invaders like bacteria or viruses. 
This task is performed by T-cells, specialised white blood cells, which are trained to distinguish foreign from self molecules and to start a specific immune response against an invader upon detection.

The T-cells face an enormous challenge since the foreign antigens are few, and they must be discerned against  a large fluctuating background of (harmless) self antigens.
They meet this challenge with the help of  a large, but restricted, number of different T-cell receptor types as well as a maturation process known as
negative selection. 

Van den Berg et al.~\cite{BRB} presented the first version of a T-cell model that takes the probabilistic nature of foreign-self discrimination into account.
They modelled the encounter of a  T-cell with an antigen-presenting cell (APC;
another type of white blood cell) in terms of sums of  i.i.d.~random variables that represent the stimulation rates emerging from the antigens displayed on the APC surface.
The crucial quantity then is the \emph{activation probability} of the T-cell, i.e., the probability that the \mbox{sum $G(z_f)$} of stimulation rates exceeds a certain activation threshold $\gact$, where $z_f$ is  the copy number of the foreign antigen type.

As T-cell activation is a rare event, analysis of biologically meaningful  activation probabilities $\mathbb{P}(G(z_f) \geq \gact)$ requires efficient simulation approaches. 
For this purpose, Lipsmeier et al.~\cite{Flo} developed an asymptotically efficient simulation approach based on  large deviation theory and taylored to this specific model.
 
Including negative selection into the model turns the crucial quantity into the \emph{conditional activation probability} $\mathbb{P}(G(z_f) \geq \gact|\Omega)$, where $\Omega$ denotes the event of survival of negative selection. 
In contrast to previous approaches, this requires individual-based T-cell modelling and  simulation.
We develop a corresponding simulation method and explore the resulting ability of foreign-self distinction. 
More precisely, we investigate the effects of two contrasting modes of 
antigen presentation during the selection process, namely so-called promiscuous presentation and presentation in tissue-specific subsets~\cite{Corinna}. 
\end{abstract}

\section{Introduction}

The reliable distinction between the body's own peptides and foreign ones is an essential prerequisite for the functioning of the immune system.
On the one hand, a wide variety of pathogens has to be identified, even if they are as yet unknown to the organism.
On the other hand, a response to the body's own molecules has to be avoided as this would result in dangerous auto-immune reactions.

The first step in this chain of events is an encounter between a T-cell and an antigen-presenting cell (APC). 
The APC displays a mixture of self and foreign antigens on its surface (a sample of the molecules around in the body); the T-cell examines the APC by means of its receptors and ultimately decides whether or not to react, i.e., to start an immune response. 
To be biologically more precise, we consider the encounters of so-called naive T-cells with professional APCs in the secondary lymphoid tissue. 
A naive T-cell is a cell that has completed its maturation process in the thymus and has been released into the body, where it has not yet been exposed to antigens. 
It tends to dwell in secondary lymphoid tissue like lymph nodes, where it comes into contact with professional APCs, special white blood cells with so-called MHC molecules at their surface that serve as carriers for antigens.

Despite the enormous challenge (one or a few foreign types of potentially dangerous foreign molecules must be discerned against a huge variety of harmless self molecules), the immune system takes the appropriate decision in the overwhelming majority of cases. 
There are two prerequisites to this ability. 
First, there is a large number (estimated at $2 \cdot 10^6$ in humans and $2 \cdot 10^7$ in mice \cite{turner_structural_2006}) of different T-cell receptors (TCRs) present in an individual. 
Each T-cell is characterised by one such receptor (which is displayed in many copies on the surface of the particular T-cell). 
For virtually every foreign invader, there is at least one receptor that fits to at least one of the derived antigens and can thus elicit an immune response. 
Second, there is a maturation process in the thymus known as negative selection, which every young T-cell must undergo before it is released into the periphery (i.e., the remaining body, except the thymus). 
During this process, the T-cell encounters numerous APCs, in an analogous way as described above, but these APCs only present self antigens. 
T-cells that react in such encounters are eliminated; the surviving ones thus have little or no self-reactivity.
According to this brief summary, the mechanism appears to be sufficiently clear -- but there are deep problems beneath the surface.

On the one hand, there is no guarantee for any antigen-receptor pair, that the antigen is either recognised or not recognised. 
In fact, the interaction between these molecules, and the resulting stimulation rates delivered to the cell, varies continuously, depending on the relevant association and dissociation rates. 
This molecular background enables a substantial crossreactivity (or promiscuity), i.e., a given T cell receptor can show significant stimulation to a large number of different antigens. 
Crossreactivity enables a large ($10^{6}-10^{8}$), but restricted repertoire of T-cell receptors to interact with an even wider range of  possible antigens ($10^{12}-10^{13}$~\cite{van_den_berg_thymic_2003}).

The picture is further complicated by the presentation of antigens in more or less arbitrary mixtures on the APC surface. 
Information about individual antigens is difficult to extract, and the T-cell faces a needle-in-a-haystack problem in the periphery, where it has to distinguish foreign antigens against a large, fluctuating self background. 

\begin{figure}[!tpb]
\floatbox[{\capbeside\thisfloatsetup{capbesideposition={right},capbesidewidth=6cm}}]{figure}[.5\textwidth]
{\caption{Schematic representation of an immunological synapse, i.e., an encounter of a T-cell with an APC. 
Every APC displays a large variety of different types of self antigens, which occur in various copy numbers. In contrast, each T-cell is characterised by a specific type of T-cell receptor (TCR), which is displayed in many identical copies on the surface of the particular T-cell.} 
}
{\includegraphics[width=.35\textwidth]{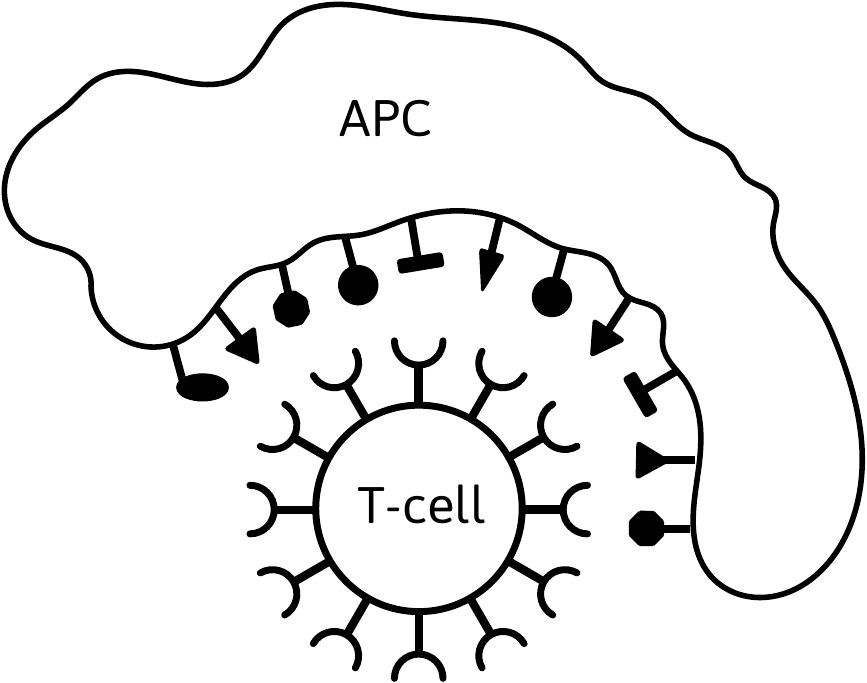}}\label{fig:01}
\end{figure}

The first approach that deals with these complications is the model by van den Berg, Rand, and Burroughs \cite{BRB}, henceforth referred to as BRB. 
This model takes into account the probabilistic nature of the problem resulting from the observations above. 
It comes in a basic version (without negative selection), which is not in itself biologically realistic but sheds light on some important aspects of foreign-self distinction in the periphery, in particular, the dependence on copy numbers of antigens (i.e., on antigen presentation); and an extended version which includes a first version of negative selection.

Knowledge of the negative selection process and, in particular, an understanding of how tolerance to tissue-restricted antigens could be achieved, is still limited.
Tissue-restricted antigens (TRAs) are derived from proteins that are only expressed in specialised tissues (as opposed to constitutive antigens that are expressed in every cell). 
In the thymus, medullary thymic epithelial cells (mTECs) are able to express TRAs in order to mediate negative selection.

Two alternative (extreme) modes of how TRAs might be presented in the thymus are under discussion~\cite{derbinski_promiscuous_2005, gillard_contrasting_2005,  klein_positive_2014}. 
According to the details of their respective genetic mechanisms, they are known as developmental model and terminal differentiation model, respectively, in the biological literature, but we would like to call them emulation model and mixture model, respectively, for the sake of the immediate intuitive appeal to the nonspecialist. 
Emulation means correlated expression that mimics particular cell lineages or tissues. 
Alternatively, under a mixture model, arbitrary antigens including TRAs would be expressed in an uncorrelated, stochastic fashion (i.e., every mTEC would express random samples of antigens, of mixed tissue origin).

The word “model” should not (yet) be understood in the sense of a mathematical model; in fact, these models have, as yet, been only formulated verbally. 
It is obvious that an emulation model can work in principle, provided the number of possible tissues or cell types is not too large and antigen presentation is emulated correctly. 
Problems may arise, however, if copy numbers fluctuate greatly. 
This is the case, for example, in secretive tissues, where certain proteins occur at times in copy numbers so large that are hard to conceive in the thymus~\cite{mason_very_1998, van_den_berg_foreignness_2004}.
On the other hand, it is entirely unclear whether mixture models can work at all. 
After all, they require that ‘dangerous’ stimulation rates be detected within their mixture – a needle-in-a-haystack problem similar to foreign recognition against a self background in the periphery.

Theoretical results on negative selection are sparse, in particular within the probabilistic framework. 
The aim of this article therefore is to explore a probabilistic model of negative selection building on the approach of BRB.
The article is organised as follows. 
In Sec.~\ref{sec:BRB}, we will review the basic BRB model (i.e., without negative selection). 
In Sec.~\ref{sec:NegSelect}, we present a mathematical formulation for the inclusion of negative selection into the BRB model. 
Our advanced model requires individual-based T-cell modelling and simulation to explore the effects of negative selection to foreign-self discrimination. 
In Sec.~\ref{sec:Sim} we develop a suitable simulation approach for that purpose. 
In Sec.~\ref{sec:Results} we examine our model of negative T-cell selection in dependence on various model parameters. 
More precisely, we explore the effects of the two contrasting modes of antigen presentation in thymus. 
It will turn out that negative selection improves the power of foreign-self discrimination due to a truncation of the tails of stimulation rate distributions.

\section{Review of basic BRB model} 
\label{sec:BRB}

\subsection{The model}

The BRB model (as introduced in~\cite{BRB} and further analysed in~\cite{Flo},\cite{hannah}, and \cite{zint}) was the first to take into account the probabilistic nature of foreign-self antigen recognition, as justified in the light of the vast variety of both T-cell receptors and possible antigens, and the problems discussed in the Introduction.
An encounter of a T-cell with an APC is modelled, where the APC presents a random mixture of antigens.
Here, we give a brief review of the basic BRB model, which does not include negative selection; the variant with negative selection will be postponed to Sect.~\ref{sec:NegSelect}.

\begin{figure}[!tpb]
\centerline{
\includegraphics[width=\textwidth]{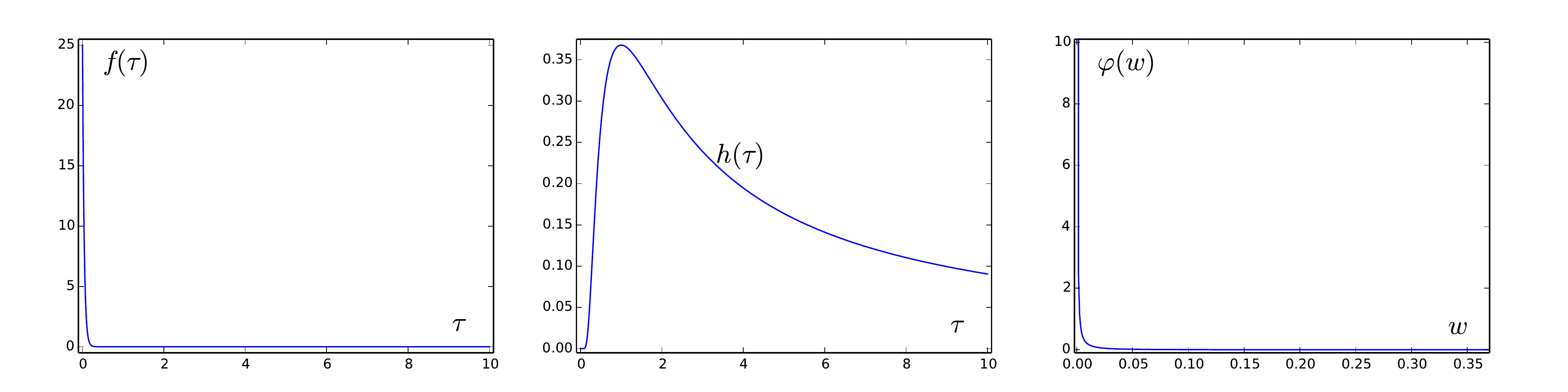}
}
\caption{Left: density of $\mathcal{T}$ ($\mathcal{T}\sim \textrm{Exp}(1/\bar{\tau})$ with $\bar{\tau} = 0.04$); middle: the function $h$ that transforms mean dwell time (random variable $\mathcal{T}$ with realisations $\tau$) into mean stimualtion rate (random variable $W$ with realisation $w$); right: density of $W=h(\Tau)$. The pole on the right (at $h(1)=1/\textrm{e} = 0.3679$) supports so little mass that it is not visible on linear scale.}\label{fig:02}
\end{figure}

\paragraph{Individual stimulation rates.}
Consider an encounter between receptor~$R_i$ and antigen~$M_j$. 
The mean dwell time (that is, the duration of a binding) of such a pair is assumed to be a random variable $\Tij$, where the $\Tij$ are i.i.d.~random variables whose distribution will be specified below. 
The mean stimulation rate emerging from this pair is $\Wij = h(\Tij)$, where $h(\tau) = \frac{1}{\tau}e^{-1/\tau}$ , that is, the dissociation rate $(1/\tau)$~times the probability~$(e^{-1/\tau})$ that the dwell time exceeds 1 (i.e., time is rescaled so that one time
unit equals the minimal binding time required for a stimulus). 
We will sometimes omit the indices of i.i.d.~random variables where appropriate, e.g., we speak of $W$ and $\Tau$ where arbitrary $\Wij$ and
$\Tij$ are meant.
In the BRB model, it is assumed that the mean dwell times $\Tij$ are i.i.d~random variables that follow the Exp($1/\bar{\tau}$) distribution, i.e., the exponential distribution with mean $\bar{\tau}$, where $\bar{\tau}$ is chosen as $\bar{\tau}=0.04$.
There is no compelling reason for the choice of exactly this distribution, and exactly this parameter, for the \textit{mean} dwell time; the choice is for convenience and for lack of knowledge of more detail. 

Fig.~\ref{fig:02} shows the function $h$, as well as the densities of $\mathcal{T}$ and the transformed random variable $W$ (these densities are denoted by $f$ and $\varphi$, respectively). 
Clearly, $f$ has practically all its mass close to 0, so that the declining part of the optimum curve $h$ is hardly ever sampled (it is, in fact, sampled so rarely that it is irrelevant). 
Consequently, $\varphi$ has most of its mass near 0 as well, but a thin tail reaches out far to the right. 
In fact, $\varphi$ has poles at 0 and at $h(1) = 1/e$, but, the right pole supports extremely little probability mass. 

\paragraph{Total stimulation rate.} 
Let us now consider the APC as a whole. 
It is assumed to present only one foreign antigen type, present in $z_f$ copies (where $z_f$ may be zero); and two classes of self antigen types (constitutive and variable), of which $n_c$ ($n_v$) are presented on every APC. 
If $z_f = 0$, they are displayed in $z_c$ ($z_v$) copies each, where $n_c<n_v$ and $z_c>z_v$.
If $z_f>0$, the self antigens are proportionally displaced, so that the total number of antigens is constant (at $M = n_c z_c + n_v z_v$), independently of $z_f$.
Let now T-cell $i$ ($T_i$) meet a random APC. 
Summing up all stimulation rates it receives from its receptors, the cell experiences the total stimulation rate
\begin{equation}
\label{formula:basicBRB}
G_i(z_f) = qz_c\left(\sum_{j=1}^{n_c}W_{ij}\right) + qz_v\left(\sum_{j=n_c+1}^{n_c+n_v}W_{ij}\right) + z_fW_{i,n_c+n_v+1},
\end{equation}
where $q = q(z_f) := (M - z_f )/M$ is the proportional displacement factor. Eq.~(\ref{formula:basicBRB}) is the \emph{basic BRB model}. 
Note that we consider $G_i$ as a function of $z_f$. 
The other parameters are fixed at $n_c = 50$, $n_v = 1500$, $z_c = 500$, and $z_v = 50$, in line with BRB. 
Together with $\bar{\tau}= 0.04$ (the mean of the exponential mean dwell time distribution), these values constitute our basic BRB parameter set, which will be used unless stated otherwise.

\paragraph{Immune response.}
It is assumed that $T_i$ starts an immune response provided $G_i(z_f)$ surpasses a threshold value $\gact$. 
The task now is to distinguish the single foreign antigen in~(\ref{formula:basicBRB}) against the
large self-background. 
This appears difficult since, by the i.i.d.~assumption on the $\Wij$ (for all $j$), there is no a-priori information about the nature of the antigen (recall that negative selection is not yet in place). 
To see whether and how recognition may work nevertheless, one considers the \emph{activation probability} $\mathbb{P}(G(z_f)\geq \gact)$, i.e., the probability of a single T-cell to become activated during an encounter. 
Assuming a total of $n$ T-cells that get in contact with APCs that present a given foreign antigen, the two crucial probabilities are
\begin{equation}
\alpha := \mathbb{P}(\textrm{at least one T-cell reacts if }z_f = 0) = 1-(1-\mathbb{P}(G(0) \geq \gact))^n
\end{equation}
and
\begin{equation}
\beta := \mathbb{P}(\textrm{no T-cell reacts if $z_f$ }> 0) = (1-\mathbb{P}(G(z_f) \geq \gact))^n .
\end{equation}
The symbols $\alpha$ and $\beta$ are chosen deliberately because of the relationship with a false positive and a false negative, respectively: $\alpha$ is the probability of an autoimmune reaction, whereas $\beta$ is the probability that a foreign antigen goes unnoticed.
We have no good knowledge of $n$ (except that it is bounded above by the number of T-cell types), but it is clear that a necessary condition for distinction is that $\gact$ can be chosen in such a way that, for physiologically realistic values of $z_f$,
\begin{equation}
\mathbb{P}(G(z_f) \geq \gact) \gg \mathbb{P}(G(0) \geq \gact).\label{eq:act_probs}
\end{equation}
There then is a region of intermediate values of $n$ for which both $\alpha$ and $\beta$ are small. 
We will, therefore, mainly investigate the validity of (\ref{eq:act_probs}) in what follows.

\subsection{Essential results for the basic BRB model}
\label{subsec:ResultsBRB}

\begin{figure}[!tpb]
\centerline{
\includegraphics[width=\textwidth]{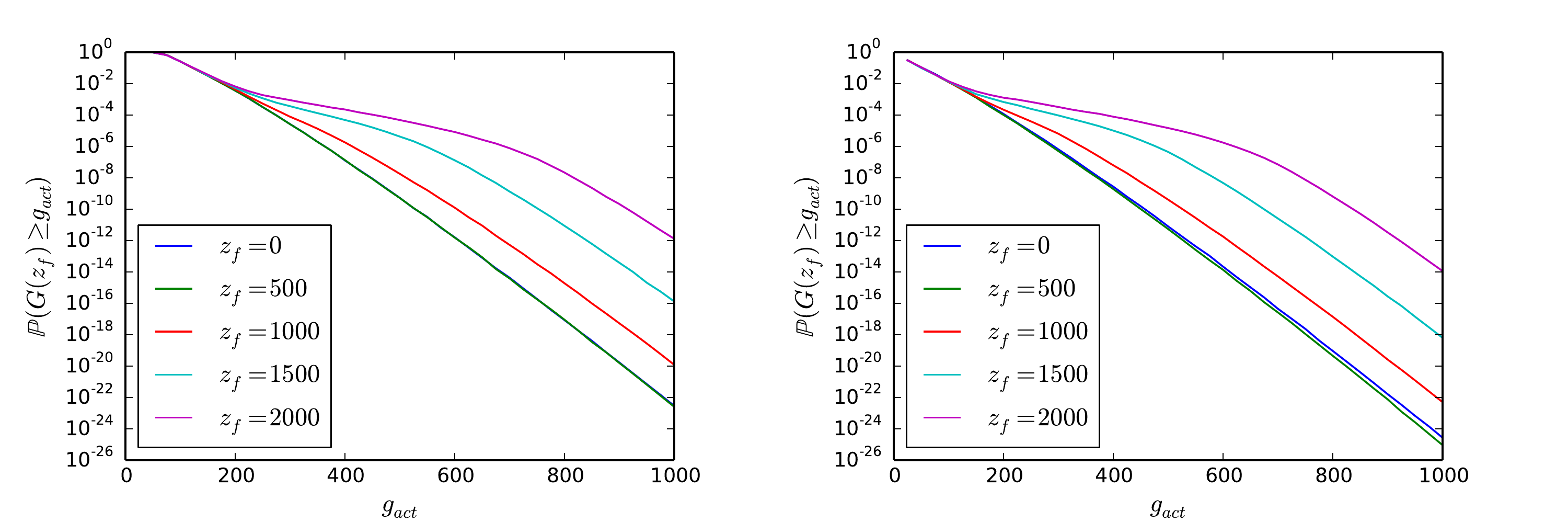}
}
\caption{Left: activation curves for the basic BRB model (\ref{formula:basicBRB}); right: activation curves in the simplified T cell model (\ref{formula:simplifiedBRB}) with $n_s=50$ and $z_s=500$.}\label{fig:03}
\end{figure}

Let us briefly summarize some previous results for the BRB model that will become essential for the understanding of negative selection.
The difficulty of the analysis lies in the small probabilities involved: Both probabilities in (\ref{eq:act_probs}) are of the order of $10^{-5}$ or less. 
Analysis of these tail events requires simulation.
The approach presented by Lipsmeier and Baake~\cite{Flo} uses large deviation theory to design an asymptotically efficient simulation method (see Sec.~\ref{sec:Sim} below).
In agreement with previous results~\cite{BRB, zint}, the authors showed that inequality~(\ref{eq:act_probs}) may be satisfied provided $z_f \gtrapprox 2z_c$, see Fig.~\ref{fig:03}.

Moreover, Lipsmeier and Baake~investigated the contributions of the constitutive and the variable stimulation rates to the self background, i.e., the contribution of the constitutive sum and the variable sum to $G(z_f)$. 
They illustrated a fundamental difference between variable and constitutive antigens. 
Variable stimulation rates are approximately normally distributed and fairly closely peaked around their mean. 
This feature even persists for subsets of samples for which $G(z_f)\geq \gact$, for various $\gact$.
So, the variable antigens form a kind of background that poses no difficulty to foreign-self distinction: it is not very noisy, and it does not change with $\gact$.
In contrast, due to the large copy numbers ($z_c=500$), the distribution of the constitutive activation rates is wider and moves to the right with increasing $\gact$.
The constitutive sum forms a fluctuating, hard-to-predict background, against which it is hard to stand out for a foreign antigen.
This demonstrates that those self antigens that are present in high copy numbers set the limit of foreign-self distinction. 

Due to these findings (many more details may be found in~\cite{Flo} and~\cite{zint}) we may restrict attention to the set of 'relevant' antigen types (that is, those that may appear in copy numbers high enough to cause problems -- a specification deliberately vague to leave room for interpretation).
In our simplified model, every APC in the periphery is assumed to present $n_s$ of these, all at the same copy number $qz_s$, where now $q = (n_sz_s-z_f)/n_sz_s$, and $z_f$ is the copy number of the one foreign antigen type
that is (maybe) also presented, as before. 
The total stimulation rate in the periphery then is
\begin{equation}
\label{formula:simplifiedBRB}
G_i(z_f) = qz_s\left(\sum_{j=1}^{n_s}W_{ij}\right) + z_fW_{i,n_s+1}.
\end{equation}
With $n_s=50, z_s=500$ (in correspondence with $n_c$, $z_c$), the activation curves of our simplified model show  qualitatively the same behaviour as for the original BRB model (see Fig.~\ref{fig:03}), i.e., (\ref{eq:act_probs}) holds for $z_f \gtrapprox 2z_c$ and sufficiently large values of $\gact$.

\section{Including negative selection into the BRB model}
\label{sec:NegSelect}

\subsection{General principles and crucial parameters}
\label{subsec:paramSet}
In order to model the process called negative selection (see~\cite{jiang_regulation_2006} for a review of the biological details), one postulates a second threshold $\gthy$ with a similar role as $\gact$. 
If the stimulation rate of a young T-cell in its maturation phase in the thymus (where it meets several APCs that only present self-peptides) exceeds this threshold in at least one out of $R$ encounters (known as \emph{rounds} of negative selection), then the T-cell is induced to die. 
The estimates of $R$ vary greatly: it may be estimated to over 2000 (from the sojourn of a young T-cell in the thymus of 4–5 days~\cite{mccaughtry_thymic_2007}, and the duration of a T-APC-meeting of 3 minutes~\cite{henrickson_t_2008}), but this contrasts with earlier findings of much lower values~\cite{muller_quantitative_2003, palmer_negative_2003}.
Altogether, a fraction $\delta$ of all T-cells is deleted during negative selection.
Estimates are usually in the range of 50\% - 65\%~\cite{stritesky_selection_2012, palmer_negative_2003}, but these values suffer from the limitations of the underlying biological models.

As to the self antigens presented, a pivotal parameter is $K$, the number of potentially immunogenic self-antigens. 
One often estimates $K = 10^4 \dots 10^5$~\cite{mason_very_1998, scherer_activationthreshold_2004}; a more recent bioinformatics analysis arrives at $2\cdot 10^5$~\cite{frankild_amino_2008}.
It is sometimes appropriate to distinguish between all $K$ self antigens and
those that may eventually appear in high copy numbers, since only these will cause problems (cf. the discussion in Sec.~\ref{subsec:ResultsBRB}). 
Let us denote their number by $\tilde{K} < K$. 
The value of $\tilde{K}$ can only be guessed; we will speculate that maybe some 10\% of all self antigens belong to this class. 
Let us now turn to our model of negative selection.

\subsection{Inclusion of negative selection into the basic BRB model}

In our (crudely simplifying) model, a round of negative selection is given by an encounter of an (immature) T cell and an APC presenting $n_s$ self antigens. 
A given T cell survives negative selection, if during $R$ rounds, in each of which a fresh APC $r$, $1 \leq r \leq R$, is presented, the total stimulation rate it perceives never exceeds $\gthy$. 
The event  `survival of negative selection' is then defined by $\Omega := \{G^{(r)}(0) < \gthy , 1 \leq r \leq R\}$.
The question now is whether or not foreign-self distinction will work on the negatively selected  T-cell repertoire in the periphery.
Hence, the crucial quantity turns into the \emph{conditional activation probability} $\mathbb{P}(G^{(R+1)}(z_f)\geq\gact|\Omega)$.

Investigation of conditional activation probabilities requires explicit individual-based T-cell modelling, as self antigens may be encountered several times during negative selection.
That is, each T-cell is given by a full set of stimulation rates. 
In our model, we restrict attention to the stimulation rates received from the $\tilde{K}$ `relevant antigens'. 
APCs present random samples of these, all at the same copy number $z_s$ (or $qz_s$ respectively). 
More precisely, our model (a first version of which appears in \cite{Flosdiss}) reads as follows: 

\begin{enumerate}
\itemsep-0.3em
\item $S := \{1,2,\dots,\tilde{K}\}$ is the set of relevant antigens.
\item T-cell $i$ is defined by its stimulation rates to \emph{all} relevant self antigens, i.e., $T_i := \{W_{i1},\dots,W_{i\tilde{K}}\}$. 
The $\Wij$ are i.i.d.~$\sim \varphi$, drawn once and fixed for the entire life of the T-cell.
\item An APC $r$ presents (and is defined by) a subset of $n_s$ of all self antigens, i.e., $A_r\subseteq S$, $|A_r|=n_s$, where the $A_r$ are all independent.
The elements of every $A_r$ are drawn from $S$ due to the presumed model of thymic antigen presentation (mixture model or emulation model respectively, cf.~Sec.~\ref{sec:APCmodels}).
One foreign antigen type can also be present, at $z_f$ copies. 
Every self antigen is displayed at the same copy number $qz_s$, $q:=(n_sz_s-z_f)/(n_sz_s)$.
\item When $T_i$ meets $A_r$, it adds together the stimulation rates it assigns to this APC's antigens, i.e., $G_i^{(r)}(z_f)=qz_s(\sum_{a\in A_r} W_{ia}) + z_fW_{n_s+1}$, where $W_{n_s+1}^{(r)} \sim \varphi$, independent of the other $\Wij$. 
\item $T_i$ survives negative selection if $G_i^{(r)}(0)<\gthy,1\leq r \leq R$.
\item A surviving T cell is activated in the periphery if $G^{(R+1)}_i(z_f)\geq \gact$.
\end{enumerate}

Note that, for notational simplicity, our definition of the $A_r$ only contains the self antigens (and no copy numbers), it is not an exhaustive description of the APCs.

\subsection{Modeling thymic antigen presentation}
\label{sec:APCmodels}

Let us now specify the sampling approach for the generation of the $A_r$, i.e., step 3 in the model above.
The mixture model argues for an arbitrary, uncorrelated TRA expression. 
Hence, under the mixture model, we assume that the  $A_r$ are sampled by drawing from $S$ independently and without replacement.

For  the emulation model, imitation of particular tissues or cell lineages is achieved by partitioning of $S$ into $s$ subsets $\check{S}_k, 1\leq k \leq s$~\cite{Corinna}. 
For simplicity, we presume subsets $\check{S},\check{S} \subset S$, to be distinct and of identical size $|S|/s$. 
Now, an APC $r$ represents a subset of $n_s$ self antigens originating preferentially from one particular subset $\check{S}_{k}$.
For each round $r$ of negative selection, $k^{(r)}$ is drawn from $\{1,2,\dots,s\}$ independently and with replacement.
Given $k^{(r)}$, a \emph{strict emulation model} scenario (in a sense that the TRAs presented by an APC originate exclusively from one particular subset) is modelled straightforwardly by drawing the elements of $A_r$ from $\check{S}_{k^{(r)}}=\{\frac{|S|}{s}(k^{(r)}-1)+1,\frac{|S|}{s}(k^{(r)}-1)+2,\dots,\frac{|S|}{s}k^{(r)}\}$ independently and without replacement. 

However, more moderate scenarios of emulation can be obtained by introducing an additional parameter $p,1/s \leq p \leq 1$, which defines the probability of any given TRA to originate from $\check{S}_{k}$. 
TRAs not originating from $\check{S}_{k}$ are drawn from $S\setminus \check{S}_{k}$ independently and without replacement.
In our model, setting $p:=1/s$ would be equivalent to the mixture model, while $p:=1$ yields a strict emulation model scenario. 
Hence, varying $p$ allows us to adjust the model continuously between the two extreme cases, with intermediate scenarios in which the TRAs in $\check{S}_{k}$ are presented preferentially, but not exclusively.

\section{Simulation approach}
\label{sec:Sim}

Our task is to simulate the conditional activation probability

\begin{equation}
\label{formula:condActProb}
\mathbb{P}(G^{(R+1)}(z_f)\geq \gact | \Omega),
\end{equation}
where $\Omega$ is the event $\Omega := \{G^{(r)}(0)<\gthy,1\leq r \leq R  \}$. 
The probability of the corresponding \emph{unconditional} event $\mathbb{P}(G^{(R+1)}(z_f) \geq \gact)$ may be simulated in the way outlined in~\cite{Flo} (after all, it is a simplified version of the basic BRB model, defined by a weighted sum of i.i.d. random variables).
More precisely, one faces the usual problem that the simple-sampling estimate,
\[
\widehat{\mathbb{P}}_\textrm{SS} (G(z_f)\geq \gact) = \frac{1}{N}\sum_{i=1}^{N}\mathbbm{1}\{G_i(z_f) \geq\gact \},
\]
where $\mathbbm{1}$ denotes the indicator function and $N$ is the number of T-cells simulated (i.e., sample size), is hopelessly inefficient because the event is so rare.
Instead, a special variant of importance sampling is employed: Denoting by $\mu$ the ‘natural’ density, one draws $\bar{G}(=\bar{G}(z_f))$ from the density $\mu^\vartheta$ obtained from $\mu$ by exponential reweighting according to
\[
{\operatorname{d}\!P\mu^\vartheta\over\operatorname{d}\!\mu}(g) = \frac{\textrm{e}^{\vartheta g}}{\mathbb{E}(\textrm{e}^{\vartheta G(z_f)})},
\]
where  $\vartheta>0$ is the \emph{tilting parameter} (to be specified below), and $\mathbb{E}(\textrm{e}^{\vartheta G(z_f)}) = \mathbb{E}_\mu(\textrm{e}^{\vartheta G(z_f)})$ is the moment-generating function of $G(z_f)$.
One then uses the importance sampling estimate 
\[
\widehat{\mathbb{P}}_\textrm{IS}(G(z_f)\geq \gact) = \frac{1}{N}\sum_{i=1}^{N}\mathbbm{1}\{\bar{G}_i(z_f) \geq\gact \}{\operatorname{d}\!\mu\over\operatorname{d}\!\mu^\vartheta}(\bar{G}_i).
\]
Here, $\operatorname{d}\!\mu/\operatorname{d}\!\mu^\vartheta(\bar{G}_i)$ serves as a reweighting factor from the `artificial' to the natural density and guarantees that the estimate is unbiased and converges to the true probability as $N \rightarrow \infty$.

The optimal choice for $\vartheta$ is well known to be the solution of 
\begin{equation}
\label{eq:tiltparam}
\mathbb{E}_{\mu^\vartheta}(\bar{G}) = {\operatorname{d}\over\operatorname{d}\!\vartheta}\log{\mathbb{E}(\textrm{e}^{\vartheta G(z_f)})} = \gact,
\end{equation}
cf.~\cite{Dieker, Flo}.
That is, the tilting is performed in such a way as to transform the rare event into a typcial one. 
For this choice of $\vartheta$, the importance sampling algorithm is known to be asymptotically efficient. 
That is, the number of samples required to keep the relative error (i.e., the standard deviation of the estimated probability divided by the true probability) below a prescribed bound increases only subexponentially with $n_s$ (if the problem is embedded into a certain sequence of problems with increasing $n_s$ and with $\gact$ scaled accordingly~\cite{Flo}), rather than exponentially as with simple sampling.

In practice, this is only useful if the underlying distribution has some further structure that may be exploited -- otherwise, the calculation of $\vartheta$ according to (\ref{eq:tiltparam}) requires the calculation of the
full distribution of $G(z_f)$, which includes the probability we want to estimate.
In the unconditional case considered so far, the independence of the summands of $G(z_f)$ (and the ensuing factorisation of the moment-generating function $\mathbb{E}(\textrm{e}^{\vartheta G(z_f)}))$ lets the problem boil down to tilting the individual $\Wij$ with parameter $qz\vartheta$ (that is, sampling new random variables $\bar{W}_{ij}$ from $\varphi^{z\vartheta}$, the tilted version of $\varphi$, where $z = qz_s$ or $z = z_f$, depending on the weighting factor the $\Wij$ is associated with). 
In line with (\ref{eq:tiltparam}), $\vartheta$ is chosen so that
\begin{equation}
\label{formula:optTheta}
\mathbb{E}_{P^\vartheta}(\bar{G}) = n_sqz_s\mathbb{E}_{\varphi^{qz_s\vartheta} }[\bar{W}] + z_f\mathbb{E}_{\varphi^{z_f\vartheta}}[\bar{W}] = \gact .
\end{equation}
Note that we use a bar throughout to indicate tilted random variables, but the individual tilting parameters do vary, as indicated by the superscript of $\varphi$.

The above is an obvious variant of the method used in~\cite{Flo}; for details, in particular on how to actually perform the tilting and the simulation in an efficient and numerically precise way, see this reference.
In practice, in the application~\cite{Flo}, the increase in computation time turned out to be only roughly linear in the number of terms in the sum, and simulation time was reduced by up to a factor of more than 1000 relative to simple sampling.

In the present context, however, we need to simulate the conditional probability in~(\ref{formula:condActProb}). 
Again, we are interested in rare events, but now we have to take care of the fact that, after negative selection, the $W_{ia}$ are not independent of a given T-cell. There is therefore no structure available that might simplify the evaluation of $\vartheta$ in the analogue of~(\ref{eq:tiltparam}), i.e., $\mathbb{E}_{\mu^\vartheta} (\overline{G^{(R+1)}|\Omega)} = \gact$ (where the overbar again indicates the tilted random variable). 
We will, in what follows, present  an unbiased importance sampling algorithm that uses a heuristic tilting scheme and is far from asymptotically efficient but makes simulations just feasible on a standard PC.

By definition, 
\begin{equation}
\label{eq:Bayes}
\mathbb{P}(G^{(R+1)}(z_f)\geq \gact | \Omega) = \frac{\mathbb{P}(\Omega_i,G^{(R+1)}(z_f)\geq \gact)}{\mathbb{P}(\Omega_i)}.
\end{equation}
We have mentioned before that $\mathbb{P}(\Omega_i) = 1-\delta$ is not small (in the sense of large deviations); but the event $\{\Omega_i , G_i^{(R+1)}(z_f) \geq \gact\}$ is even rarer than $\{G_i^{(R+1)}(z_f) \geq \gact\}$ for the relevant (large) $\gact$. 
There is no obvious importance sampling strategy for the latter probability,  but we propose to adapt the tilting scheme that would apply if the two events were independent, in which case $\mathbb{P}(G^{(R+1)}(z_f)\geq \gact | \Omega) = \mathbb{P}(G^{(R+1)}(z_f)\geq \gact)\mathbb{P}(\Omega)$.

$\mathbb{P}(G^{(R+1)}(z_f)\geq \gact)$ can be handled in the following way. 
Note that, since the $\Wij$ that define $T_i$ are i.i.d., $G_i^{(R+1)}$ is invariant under any permutation of the elements of $S$ (for all $r$), so that we may assume that $A_{R+1} = \{1,\dots,n_s\}$ without loss of generality.
We may thus efficiently simulate $\mathbb{P}(G^{(R+1)}(z_f)\geq \gact)$ via importance sampling, by sampling $\bar{W}_{ia}, 1 \leq a \leq n_s$, from $\varphi^{qz_s\vartheta}$, and (if $z_f>0$), the foreign antigen $\bar{W}_{i,\tilde{K}+1}$ from $\varphi^{z_f\vartheta}$, with $\vartheta$ determined in the usual way~(\ref{formula:optTheta}), while leaving the remaining $\Wij, ns + 1 \leq j \leq \tilde{K}$, unchanged. 
We therefore propose the following heuristic procedure:

\begin{enumerate}
\itemsep-0.3em
\item $\vartheta :=$ solution of~(\ref{formula:optTheta});
\item for T-cell $i$, sample the `partially tilted' T-cell as $T_i^* := (\bar{W}_{i1},\dots,\bar{W}_{in_s},W_{i,n_s+1},\dots,W_{i,\tilde{K}})$, \mbox{$W_{ij} \textrm{ i.i.d}\sim \varphi^{qz_s\vartheta}$};
\item if $z_f>0$, also sample $\bar{W}_{i,\tilde{K}+1}\sim\varphi^{z_f\vartheta}$;
\item sample $A_1,\dots,A_R$ independently and without replacement from $S$, due to the assumed model of thymic antigen expression (cf.~Sec.~\ref{sec:APCmodels});
\item $A_{R+1} := \{1,\dots,n_s \}$;
\item $T_i^*$ becomes activated by an encounter with $A_{R+1}$ if 
\[ 
(G_i^{(R+1)}(z_f))^* := qz_s\left(\sum_{a \in A_{R+1}} T^*_{ia} \right) + z_f\bar{W}_{i,\tilde{K}+1} \geq \gact,
\]
where $T_{ia}^*$ is the $a$-th component of $T^*_i$, i.e., $\bar{W}_{ia}$ or $W_{ia}$, respectively.
\item If $T^*_i$ is activated, check whether it survives negative selection, i.e., whether $(G_i^{(r)}(0))^* = \sum\nolimits_{a\in A_r} T^*_{ia} < \gthy$ for $1\leq r \leq R$;
\item according to~(\ref{eq:Bayes}), estimate 
\[
\begin{split}
\widehat{\mathbb{P}}_\textrm{IS}(G^{(R+1)}(z_f) \geq \gact\ |\ \Omega) =  & \frac{1}{\mathbb{P}(\Omega)}\frac{1}{N}\sum_{i=1}^N \mathbbm{1}\{ \Omega_i, (G_i^{(r)}(z_f))^* \geq \gact \} \\ 
& \times \frac{\mathbb{E}(\textrm{e}^{z_f\vartheta\bar{W}_{i,n_s+1}})}{\textrm{e}^{z_f\vartheta\bar{W}_{i,n_s+1}}}\prod_{j=1}^n\frac{\mathbb{E}(\textrm{e}^{z_s\vartheta\bar{W}_{i,j}})}{\textrm{e}^{z_s\vartheta\bar{W}_{i,j}}}
\end{split}
\]
with $\mathbb{P}(\Omega) \equiv 1-\delta$.
\end{enumerate}
 
This would be an asymptotically efficient method altogether if $A_R \cap A_{R+1} = \emptyset$ for $1\leq r \leq R$, but this is clearly not the case -- after all, it is the very nature of negative selection that (most) self antigens have been seen in the thymus.
But this, in turn, has its side effect on $\mathbb{P}(\Omega_i,G_i^{(R+1)}(z_f) \geq \gact)=\mathbb{P}(G_i^{(R+1)}(z_f) \geq \gact)\mathbb{P}(\Omega_i|G_i^{(R+1)}(z_f) \geq \gact)$, which becomes very small due to the second factor, even in the tilted version. Nevertheless, this heuristics does allow simulation, although
a single activation curve takes a few days -- but simple sampling is forbidding (on a PC, at least).

\section{Results}
\label{sec:Results}

As a basic set of model parameters we specify $\varphi = \textrm{Expo}(1/\bar{\tau}), \bar{\tau}=0.04$, and $n_s=50, z_s=500$ as in the simplified T-cell model introduced in Sec.~\ref{subsec:ResultsBRB}. 
Furthermore, we define $\tilde{K} := 1000$ (assuming that there are $10^4$ immunogenic antigens altogether (a lower limit), of which 10\% might be in high frequency (our ad hoc guess)). 
For analysis of our emulation model of thymic antigen expression the additional parameter $s$ (c.f.~Sec.~\ref{sec:APCmodels}) is required, with 
$|S|/s \geq n_s$. 
Under these requirements, $s:=S/n_s=20$ provides the highest possible number of distinct antigen subsets $\check{S}_k, 1\leq k \leq s$,  and seems a reasonable choice to begin with.
Finally, we set $\delta := 0.5$ and $R := 2000$ (which is presumably rather
at the upper end,  c.f.~Sec.~\ref{subsec:paramSet}).
For any parameter set, in the first step of simulation, $\gthy$ must be chosen so that a  given fraction $\delta$ of T-cells is deleted.

\subsection{The action of negative selection}

Negative selection alone (i.e., without any APC encounters in the periphery) is straightforward to simulate via simple sampling since neither survival nor death of cells are rare events (in the sense of large deviation theory).

\begin{figure}[!tpb]
\centerline{
\includegraphics[width=\textwidth]{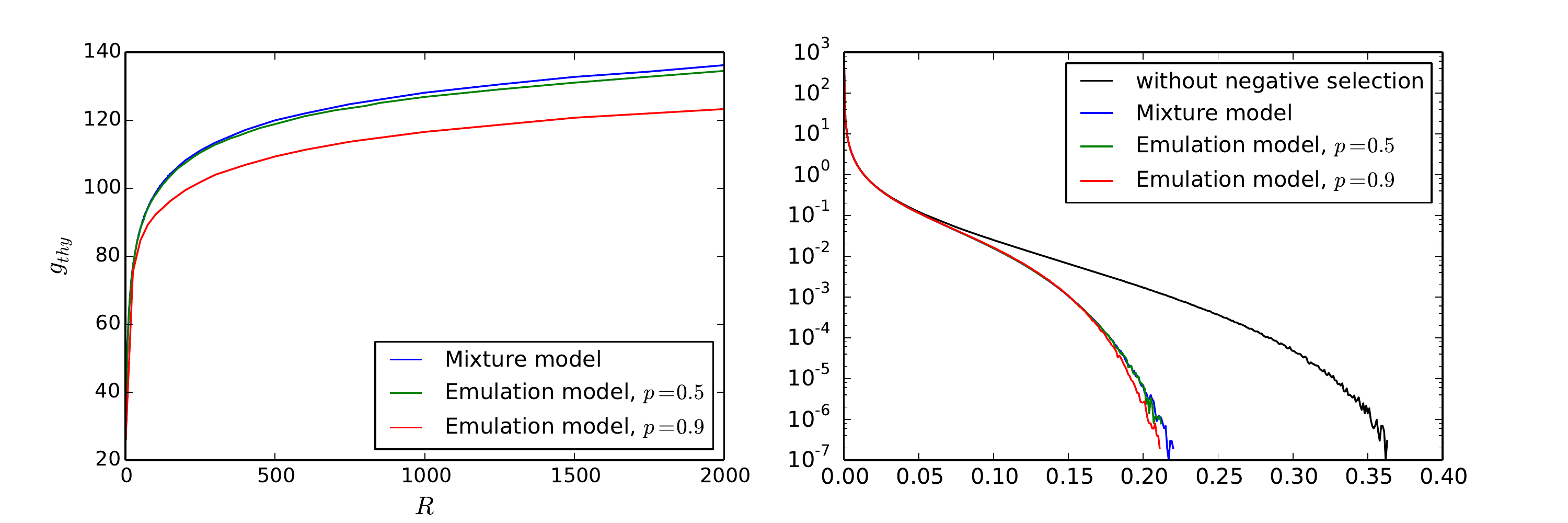}
}
\caption{Left: The thymic threshold $\gthy$ required to eliminate a fraction of $\delta=0.5$ of T-cells in $R$ rounds of negative selection; right: Original empirical density of stimulation rates and empirical densities of stimulation rates received from T-cells surviving negative selection.}\label{fig:05}
\end{figure}

\paragraph{Thymic threshold as a function of $R$.} 
Fig.~\ref{fig:05} (left) shows the dependence of $\gthy$ on $R$ and the presumed model of thymic antigen presentation for our basic parameter set. 
A sample size of $10^6$ cells was simulated each time, for the mixture model, and for the emulation model  with $p=0.5$ and $p=0.9$. 
We renounce simulation of a strict emulation model scenario with $p=1$, since with our basic parameter set, the sets of TRAs presented by an APC would be identical to subsets $\check{S}_k$.
Hence, for physiologically realistic values of $\gact$, activation of T cells by self antigens only, i.e., $z_f=0$, would only be possible if one subest $\check{S}_k$ of antigens were never  presented during $R$ rounds in the thymus, but were presented later in the periphery. 

Due to the independence of stimulation rates received from a single APC, for $R=1$ round of negative selection the  model of thymic antigen presentation has no effects on $\gthy$. 
But, the influence of the mode of TRA expression on thymic threshold $\gthy$ increases with $R$. 
Thereby, the values of $\gthy$ decrease with increasing strictness of the underlying emulation model, i.e., increasing $p$. 
(Note that, in our formulation the mixture model of thymic antigen presentation is equivalent to an emulation model scenario with $p:=1/s$.) 
With $R=2000$ an immature T-cell has seen each relevant antigen with high probability (exact values rely on the mode of antigen expression, e.g.~$1-(1-n_s/\tilde{K})^R=1-2.8\cdot10^{-45}$ for the mixture model), but it has not seen them in all possible combinations, which is why saturation is not yet complete.

\paragraph{Empirical posterior distribution.} Negative selection will act on activation probabilities in two ways: It will change the (marginal) stimulation rate distribution (for example, by cutting away part of its tail), and it will introduce correlations, e.g., by mutilating the simultaneous occurrence of two or more intermediate stimulation rates on any one T-cell. 
We would like to investigate the relative importance of these two contributions.
To this end, let $\check{\varphi}$ be the density of $W$ conditional on $\Omega$.
We estimate it via the collection $\{W_{ij}\}_{i\in I, j \in S}$, where $I$ is the surviving set $I := \{i \in \{ 1,2,\dots, N \}\ |\ \Omega_i \}$.
The corresponding empirical density is an estimate of $\check{\varphi}$; by slight abuse of notation, we will denote this empirical version by $\check{\varphi}$ as well.

Fig.~\ref{fig:05} (right) shows that negative selection indeed has a profound effect on the distribution of the stimulation rates in that it cuts away a significant part of the density’s tail (roughly half of its length for $\bar{\tau} = 0.04$). 
This  reduction of the `self background' by negative selection is practically independent of the  mode of antigen expression.
However, truncation of the right tails of $\check{\varphi}$ seems to increase very slightly with the strictness of emulation, i.e., with increasing $p, 1/s \leq p \leq 1$.

\subsection{Activation curves}

\begin{figure}
\begin{subfigure}{\linewidth}
\centering
\includegraphics[width=.5\textwidth]{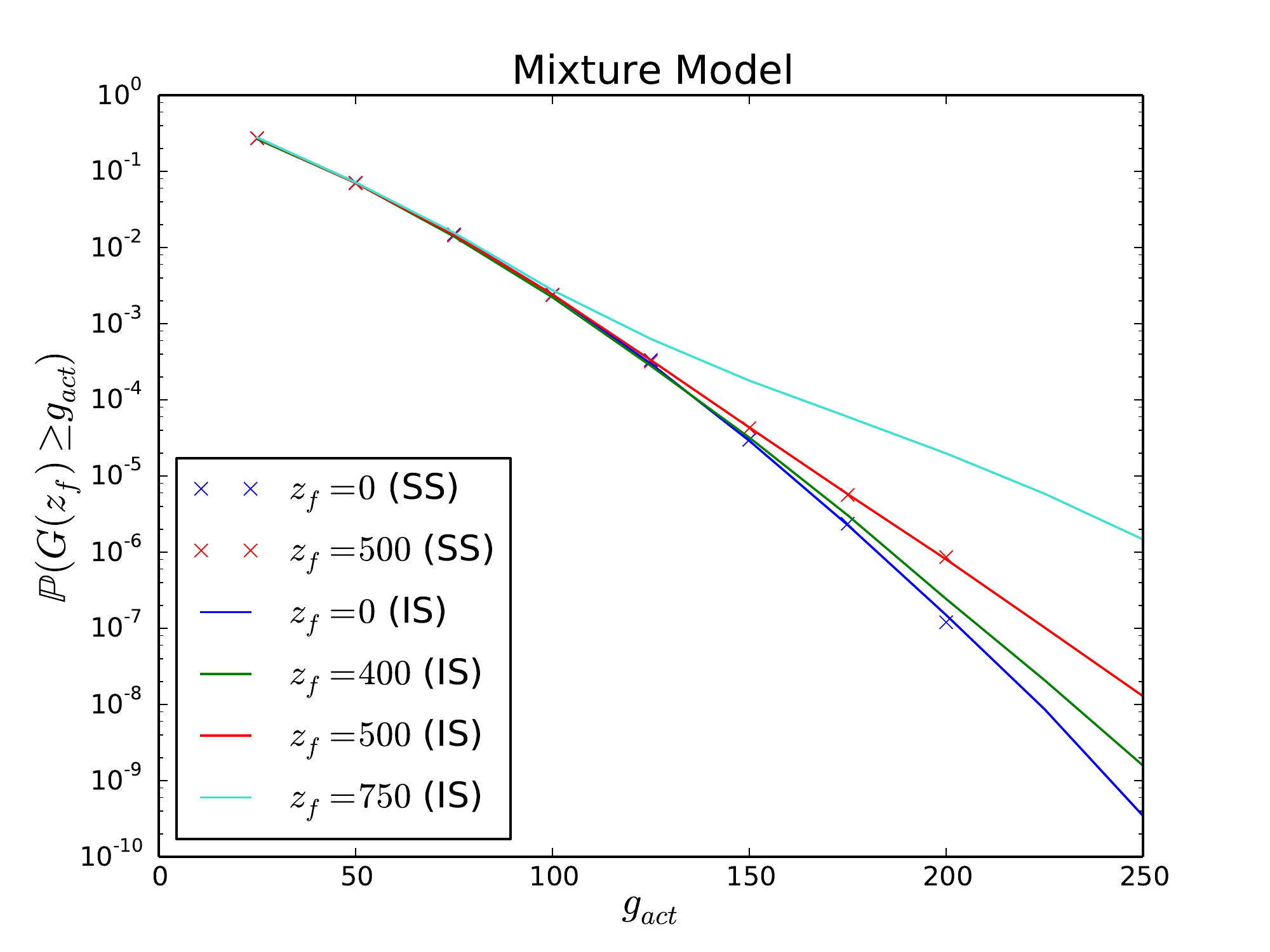}
\caption{}
\label{fig:sub_mixture}
\end{subfigure}
\begin{subfigure}{.5\linewidth}
\centering
\includegraphics[width=\textwidth]{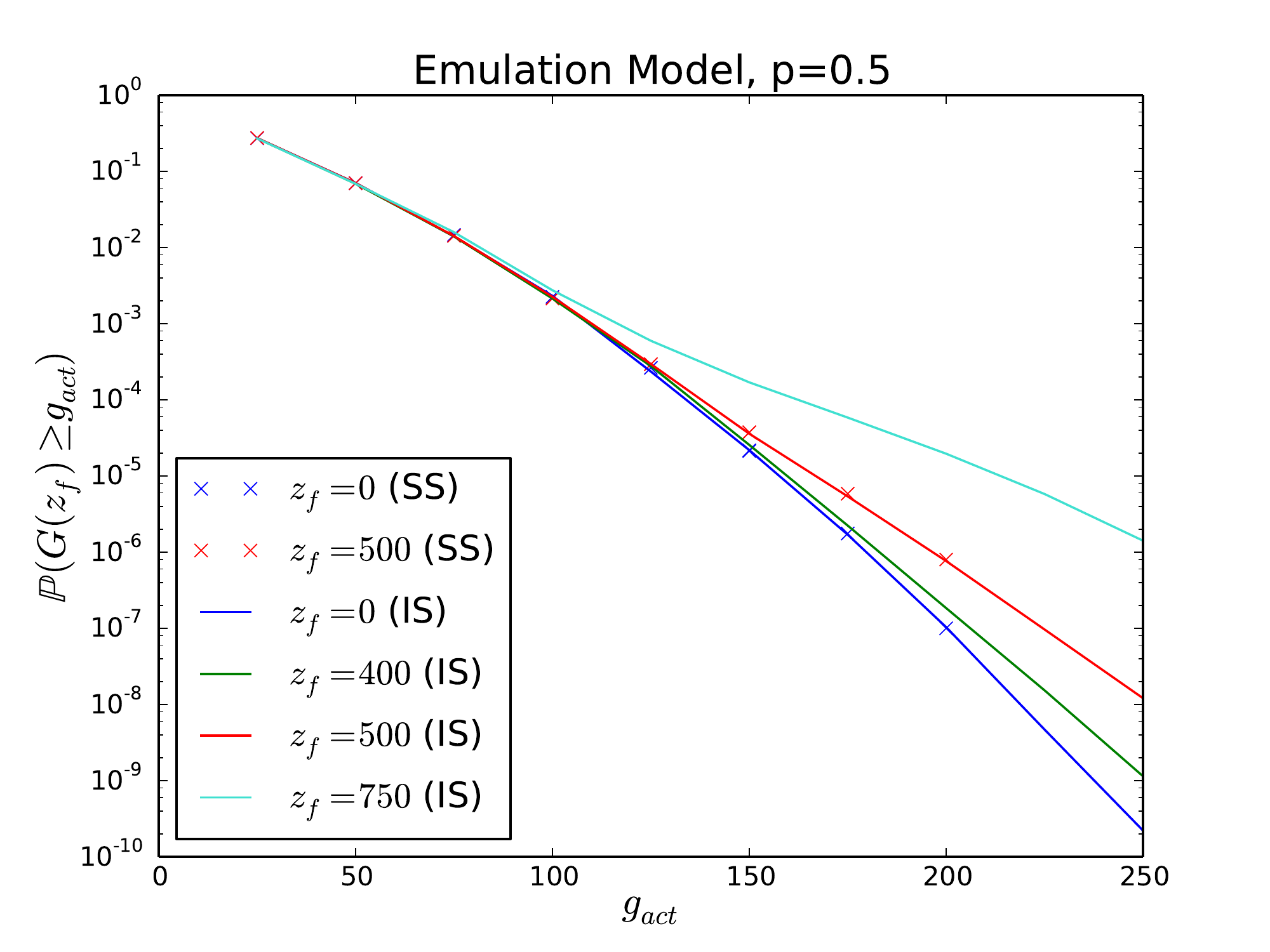}
\caption{}
\label{fig:sub_emu05}
\end{subfigure}%
\begin{subfigure}{.5\linewidth}
\centering
\includegraphics[width=\textwidth]{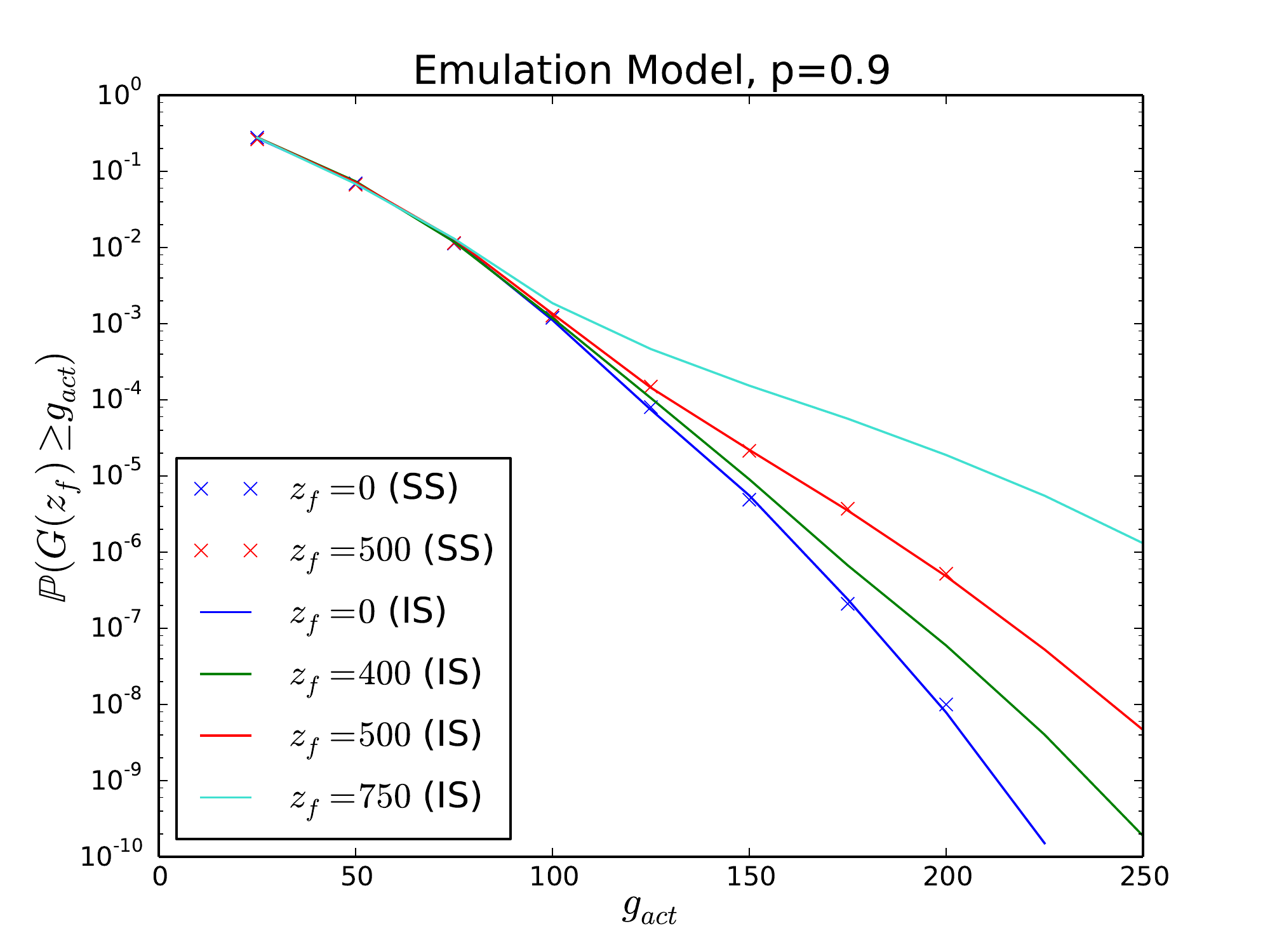}
\caption{}
\label{fig:sub_emu09}
\end{subfigure}\\

\caption{Activation curves for negative selection via mixture model (Fig.~\ref{fig:sub_mixture}), emulation model with $p=0.5$ (Fig.~\ref{fig:sub_emu05}) and emulation model with $p=0.9$ (Fig.~\ref{fig:sub_emu09}).}
\label{fig:06}
\end{figure}

Fig.~\ref{fig:06} shows the activation curves for our basic parameter set and our collection of antigen presentation modes.
Data were generated via the simulation approach introduced in Sec.~\ref{sec:Sim}. 
In each case, sample size was arranged so that  $\hat{\sigma}_\textrm{IS} \leq 10^{-3}\hat{\mathbb{P}}_\textrm{IS}(G^{(R+1)}(z_F)\geq g_{act}\ |\ \Omega)$, where  $\hat{\sigma}_\textrm{IS}$ is the standard deviation of the sampled estimators (and $10^{-3}$ is an ad hoc choice).

To validate  our simulation approach, we also ran    simple sampling  for $z_f=0$ and $z_f=z_s=500$ and  sample size  $N=10^8$.
Estimates of the conditional probabilities obtained from  simple sampling and importance sampling agree well -- at least for the small values of $\gact$ feasible with simple sampling (cf.~Fig.~\ref{fig:06}), i.e., $\gact\leq 200$.

Irrespective of the  mode of  antigen presentation, negative selection shows its efficiency in the separation of activation curves for sufficiently large values of $\gact$ and $z_f \geq z_s$.
In contrast, the activation curves for the simplified T-cell model without negative selection require values of $z_f \geq 2z_s$ (cf.~Fig.~\ref{fig:03}).
In case of the stringent emulation model  with $p=0.9$, curves even separate in the presence of a foreign antigen presented in a copy number slightly less than the copy number of self antigens, i.e., $z_s=400$. 
This might be a hint that -- at least in our theoretical framework -- the emulation model of thymic antigen presentation is more effective than the presentation of self antigens in arbitrary mixtures.
This may be related to the slightly increased truncation of the empirical post-selection density (cf.~Fig.~\ref{fig:05}). 
However, it is more plausible that the correlations between the individual stimulation rates differ markedly between the two models, and are more important than the (marginal) density as such; this remains to be investigated. Also, 
the parameter space must be explored further, e.g.~with higher $\tilde{K}$, to  simulate a strict emulation model scenario with $p=1$.

\section{Conclusion}

We have presented a probabilistic T-cell model that includes negative selection and takes  contrasting models of TRA expression in the thymus into account.
Inclusion of negative selection into the basic BRB model required individual-based T-cell modelling, which does not lend itself to the asymptotically effcient IS approach introduced by Lipsmeier and Baake~\cite{Flo}.
However, we presented a simulation approach based on `partial tilting' of the stimulation rates recognized by a single T-cell.
Although our approach is far  from being asymptotically efficient, it allowed investigation of the effects of negative selection by a pilot simulation for diverging modes of thymic antigen presentation, namely arbitrary TRA presentation, and more or less strict emulation of tissue-specific cell lines.
 
We observed that negative selection leads to truncation of the tail  of the distribution of the stimulation rates mature T-cells receive from self antigens, i.e., the  self background is reduced. 
This increases the activation probabilities of single T-cells in the presence of non-self antigens presented in copy numbers identical to those of self antigens. 

\subsubsection*{Acknowledgement}
It is our pleasure to thank Florian Lipsmeier, Jens Derbinski, and Bruno Kyewski for stimulating discussions in the early phase of the project.
We thank Sven Rahmann for providing computational infrastructure.

\small

\begin{thebibliography}{10}

\bibitem{derbinski_promiscuous_2005}
J.~Derbinski, J.~G{\"a}bler, B.~Brors, S.~Tierling, S.~Jonnakuty,
  M.~Hergenhahn, L.~Peltonen, J.~Walter, and B.~Kyewski.
\newblock Promiscuous gene expression in thymic epithelial cells is regulated
  at multiple levels.
\newblock {\em J Exp Med}, 202:33--45, 2005.

\bibitem{Dieker}
A.~B. Dieker and M.~Mandjes.
\newblock On asymptotically efficient simulation of large deviation
  probabilities.
\newblock {\em Adv Appl Probab}, 37:539--552, 2005.

\bibitem{Corinna}
C.~Ernst.
\newblock Simulating contrasting models of thymic selection via importance
  sampling.
\newblock Master's thesis, Faculty of Technology, Bielefeld University, 2011.

\bibitem{frankild_amino_2008}
S.~Frankild, R.~J. de~Boer, O.~Lund, M.~Nielsen, and C.~Kesmir.
\newblock Amino acid similarity accounts for {T} cell cross-reactivity and for
  {“Holes”} in the {T} cell repertoire.
\newblock {\em {PLoS} {ONE}}, 3:e1831, 2008.

\bibitem{gillard_contrasting_2005}
G.~O. Gillard and A.~G. Farr.
\newblock Contrasting models of promiscuous gene expression by thymic
  epithelium.
\newblock {\em J Exp Med}, 202:15--19, 2005.

\bibitem{henrickson_t_2008}
S.~E. Henrickson, T.~R. Mempel, I.~B. Mazo, B.~Liu, M.~N. Artyomov, H~. Zheng,
  A.~Peixoto, M~.P. Flynn, B.~Senman, T.~Junt, H.~C. Wong, A.~K. Chakraborty,
  and von U.~H.~Andrian.
\newblock T cell sensing of antigen dose governs interactive behavior with
  dendritic cells and sets a threshold for {T} cell activation.
\newblock {\em Nat Immunol}, 9:282--291, 2008.

\bibitem{jiang_regulation_2006}
H.~Jiang and L.~Chess.
\newblock Regulation of immune responses by {T} cells.
\newblock {\em New Engl J Med}, 354:1166--1176, 2006.

\bibitem{klein_positive_2014}
L.~Klein, B.~Kyewski, P.~M. Allen, and K.~A. Hogquist.
\newblock Positive and negative selection of the {T} cell repertoire: what
  thymocytes see (and don't see).
\newblock {\em Nat Rev Immunol}, 14:377--391, 2014.

\bibitem{Flosdiss}
F.~Lipsmeier.
\newblock {\em Rare event simulation for probabilistic models of {T}-cell
  activation}.
\newblock PhD thesis, Faculty of Technology, Bielefeld University, 2010.

\bibitem{Flo}
F.~Lipsmeier and E.~Baake.
\newblock Rare event simulation for {T}-cell activation.
\newblock {\em J Stat Phys}, 134:537--566, 2009.

\bibitem{mason_very_1998}
D.~Mason.
\newblock A very high level of crossreactivity is an essential feature of the
  {T}-cell receptor.
\newblock {\em Immunol Today}, 19:395--404, 1998.

\bibitem{hannah}
H.~Mayer and A.~Bovier.
\newblock Stochastic modelling of {T}-cell activation.
\newblock {\em J Math Biol}, 2014 (online first, doi
  10.1007/s00285-014-0759-x).

\bibitem{mccaughtry_thymic_2007}
T.~M. McCaughtry, M.~S. Wilken, and K.~A. Hogquist.
\newblock Thymic emigration revisited.
\newblock {\em J Exp Med}, 204:2513--2520, 2007.

\bibitem{muller_quantitative_2003}
V.~M\"uller and S.~Bonhoeffer.
\newblock Quantitative constraints on the scope of negative selection.
\newblock {\em Trends Immunol}, 24:132--135, 2003.

\bibitem{palmer_negative_2003}
E.~Palmer.
\newblock Negative selection — clearing out the bad apples from the {T}-cell
  repertoire.
\newblock {\em Nat Rev Immunol}, 3:383--391, 2003.

\bibitem{scherer_activationthreshold_2004}
A.~Scherer, A.~Noest, and R.~J. de~Boer.
\newblock Activation–threshold tuning in an affinity model for the {T}–cell
  repertoire.
\newblock {\em Proc Biol Sci}, 271:609--616, 2004.

\bibitem{stritesky_selection_2012}
G.~L. Stritesky, S.~C. Jameson, and K.~A. Hogquist.
\newblock Selection of self-reactive {T} cells in the thymus.
\newblock {\em Annu Rev Immunol}, 30:95--114, 2012.

\bibitem{turner_structural_2006}
S.~J. Turner, P.~C. Doherty, J.~{McCluskey}, and J.~Rossjohn.
\newblock Structural determinants of {T}-cell receptor bias in immunity.
\newblock {\em Nat Rev Immunol}, 6:883--894, 2006.

\bibitem{van_den_berg_thymic_2003}
H.~A. van~den Berg and C.~Molina-Par{\'i}s.
\newblock Thymic presentation of autoantigens and the efficiency of negative
  selection.
\newblock {\em Comput Math Method M}, 5:1--22, 2003.

\bibitem{van_den_berg_foreignness_2004}
H.~A. van~den Berg and D.~A. Rand.
\newblock Foreignness as a matter of degree: the relative immunogenicity of
  {peptide/MHC} ligands.
\newblock {\em J Theor Biol}, 231:535--548, 2004.

\bibitem{BRB}
H.~A. van~den Berg, D.~A. Rand, and N.~J. Burroughs.
\newblock A reliable and safe {T} cell repertoire based on low-affinity {T}
  cell receptors.
\newblock {\em J Theor Biol}, 209:465--486, 2001.

\bibitem{zint}
N.~Zint, E.~Baake, and F.~den Hollander.
\newblock How {T}-cells use large deviations to recognize foreign antigens.
\newblock {\em J Math Biol}, 57:841--861, 2008.

\end{thebibliography}
\bibliographystyle{plain}

\end{document}